\documentclass[12pt]{article}
\usepackage{graphicx,color}
\begin{document}
\baselineskip=5.5mm
\newcommand{\be} {\begin{equation}}
\newcommand{\ee} {\end{equation}}
\newcommand{\Be} {\begin{eqnarray}}
\newcommand{\Ee} {\end{eqnarray}}
\renewcommand{\thefootnote}{\fnsymbol{footnote}}
\def\a{\alpha}
\def\b{\beta}
\def\g{\gamma}
\def\G{\Gamma}
\def\d{\delta}
\def\D{\Delta}
\def\e{\epsilon}
\def\k{\kappa}
\def\l{\lambda}
\def\L{\Lambda}
\def\t{\tau}
\def\om{\omega}
\def\Om{\Omega}
\def\s{\sigma}
\def\lg{\langle}
\def\rg{\rangle}
\newcommand{\tblue}[1]{\textcolor{blue}{#1}}
\begin{center}
{\large {\bf Higher-order correlation functions and nonlinear response functions in a Gaussian trap model} }\\
\vspace{0.5cm}
\noindent
{\bf Gregor Diezemann} \\
{\it
Institut f\"ur Physikalische Chemie, Universit\"at Mainz,
Duesbergweg 10-14,\\ 55128 Mainz, FRG\\
}
\end{center}
PACS: 64.70.P-, 64.70.Q-, 61.20.Lc, 05.40.-a\\

\noindent
{\it
The four-time correlation function of a general dynamical variable obeying Gaussian statistics is calculated for the trap model with a Gaussian density of states.
It is argued that for energy-independent variables this function is reminiscent of the four-time functions that have been discussed earlier in the interpretation of the results of four-dimensional NMR experiments on supercooled liquids.
Using an approximative relation between the four-time correlation function and the cubic response function
the nonlinear susceptibility is calculated and the results are compared with the corresponding ones resulting from an exact calculation.
It is found that the results of the approximation change the qualitative behavior of the modulus of the susceptibility.
Whereas in the exact calculation a peak is found in the modulus in most cases, depending on temperature and the additional model parameters no such peak occurs in the approximation.
This difference has its origin mainly in an incorrect estimate of the static response.
The results are discussed in relation to recent experimental findings.
}

%
\section*{I. Introduction}
In the last decade a number of steps have been undertaken in order to deepen our understanding of the heterogeneous dynamics of supercooled liquids and 
glasses\cite{Berthier:2011p6852,G23,Sillescu99,Ediger00,Richert02}.
In many investigations certain higher-order time correlation functions play an important role. This holds for computer simulation studies\cite{Schroder03, Reichman07}, frequency-selective techniques like nonresonant 
holeburning\cite{SBLC96, G16, G23} and four-dimensional NMR experiments\cite{SRS91, HWZS95, G13}.
Apart from these methods to characterize the dynamic heterogeneities it has been shown how a length scale can be  extracted from a special four-point correlation function $\chi_4(t)$\cite{Berthier05} and its detailed properties have been studied theoretically\cite{Toninelli05, Berthier07a, Berthier07b}.
It is to be noted that the special function $\chi_4(t)$ differs from the corresponding four-time correlation function that had been used earlier in the experimental determinations of the length scale associated with the heterogeneities of supercooled liquids\cite{Tracht98, Reinsberg01}.

Since it is not straightforward to investigate $\chi_4(t)$ experimentally, it is necessary to have observables at hand that are related to it in a clearcut manner.
In order to provide a means for the experimental determination of the number of correlated particles, 
$N_{\rm corr}$, or the length scale, Bouchaud and Biroli related the nonlinear (cubic) response $\chi_3(\om,T)$ to a four-point function\cite{Bouchaud05}.
This relation was utilized in the experimental determination of $N_{\rm corr}$\cite{CrausteThibierge10, Brun11}.
It was argued that the modulus of the cubic response function, $|\chi_3(\om,T)|$, exhibits a hump-like structure which is assumed to be a distinctive feature of glassy correlations\cite{CrausteThibierge10}.
The maximum of $|\chi_3(\om,T)|$ is expected to decrease with increasing temperature and to be directly proportional to $N_{\rm corr}$.
In systems without glassy correlations one does not expect a peak but a 'trivial' behavior, i.e. a smooth decay to zero starting from a finite low-frequency limit as a function of frequency.

Recently, nonlinear dielectric experiments have also been performed in a different context to investigate the dynamic heterogenenities of supercooled liquids\cite{Richert06, Weinstein07} and to measure the configurational heat capacity of liquids\cite{Wang07}.

A nonlinear response theory for Markov processes has been presented in ref.\cite{G75}, to be denoted as I in the following.
The general theory was applied to two simple models for glassy relaxation, the model of dipole reorientations in an asymmetric double well 
potential (ADWP-model)\cite{Frohlich49, G46} and the well-studied trap model with a Gaussian density of states\cite{Dyre95, MB96, Denny03, G64, G71, G73}. 
The results of the model calculations suggest that a direct relation between the cubic response function and some type of glassy correlations cannot be shown to exist in general.
For the ADWP-model, a peak in the modulus of the cubic response is observed only in a certain narrow temperature range and only for non-vanishing asymmetry.
For most temperatures the model exhibits 'trivial' behavior. 
The trap model shows both, a peak or trivial behavior, depending on the variable chosen and on temperature. 
Furthermore, for some specific choice of the dynamical variable used to probe the dynamics, the peak-maximum increases as a function of temperature and for other choices it decreases.
Both models are of a mean-field type and therefore of course do not show any aspects of spatial correlations as they exist in finite dimensional systems.

The general theory provides expressions for the nonlinear response functions that cannot be related to well-defined time correlation functions in any obvious manner.
In the theoretical paper by Bouchaud and Biroli\cite{Bouchaud05}, it was argued that for systems obeying a Langevin dynamics, the cubic response near a phase transition should be related to the four-time correlation function via an expression that reminds of a fluctuation-dissipation theorem (FDT):
\be\label{R3.FDT}
R_3(t_0,t_1,t_2,t_3)\sim\b^3{d^3\over dt_1dt_2dt_3}\lg M(t)M(t_1)M(t_2)M(t_3)\rg
\ee
Here, $M(t)$ is the dynamical variable conjugate to the applied field, $t_0>t_1>t_2>t_3$, $R_3(t_0,t_1,t_2,t_3)$ is the (impulse) response to field kicks taking place at $t_1$, $t_2$ and $t_3$ and $\b=1/T$ with the Boltzmann constant set to unity.
This relation is meant to indicate that the right hand side is the dominant contribution to the response and that no other term contributing to it will be more divergent.

In the present paper, I will calculate the four-time correlation function occuring in eq.(\ref{R3.FDT}), investigate its relation to the response and furhermore show that for the special case of a trap model this function is directly related to the class of four-time functions observed in four-dimensional NMR.
In the next Section, this relation will be discussed in more detail and afterwards the cubic response originating from eq.(\ref{R3.FDT}) will be compared to the exact expressions given in ref.\cite{G75}.
Finally, some conclusions will close the paper.
\section*{II. Four-time correlation functions for trap models}
In order to calculate correlation functions of a dynamical variable $M(t)$ obeying a Markovian dynamics one uses the relevant probability functions as obtained from the dynamical rules.
The stochastic dynamics for the trap model is defined by the master equation (ME) for the conditional probability to find the system in the trap characterized by the trap energy $\e$ at time $t$ provided it was in trap $\e_0$ at $t_0$, $G(\e,t+t_0|\e_0,t_0)=G(\e,t|\e_0,0)\equiv G(\e,t|\e_0)$:
\be\label{ME.G}
{\dot G}(\e,t|\e_0)= -\k(\e)G(\e,t|\e_0)+\rho(\e)\!\int\!d\e'\k(\e')G(\e',t|\e_0) 
\ee
In eq.(\ref{ME.G}), the escape rate is given by
\be\label{k.T}
\k(\e)=\k_\infty e^{\b\e}
\ee
with the attempt rate $\k_\infty$.
The model with a Gaussian density of states (DOS) is defined by
\be\label{DOS.Gauss}
\rho(\e)\!=\!{1\over\sqrt{2\pi}\s}e^{-\e^2/(2\s^2)}
\ee
with $\s=1$. 
Note that in this model the system reaches equilibrium at all temperatures $T$ (measured in units of $\s$) and the equilibrium populations are found to be Gaussian 
$p^{\rm eq}(\e)=\lim_{t\to\infty}G(\e,t|\e_0)={1\over\sqrt{2\pi}\s}e^{-(\e-{\bar\e})^2/(2\s^2)}$
with ${\bar\e}=-\b \s^2$.
In contrast, the often studied model with an exponential DOS shows a transition to a low-temperature phase in which no equilibrium is reached and the system ages for all times\cite{MB96}.
In the following calculations I will always assume that the system is in thermal equilibrium. Thus aging effects are unimportant in the present context.

The two-time correlation function (2t-CF) of a variable $M(t)$ in general is given by:
\be\label{C2.Pi}
C_M(t,t_0)=\lg M(t)M(t_0)\rg
=\int\!d\e\int\!d\e_0 M(\e)M(\e_0)G(\e,t-t_0|\e_0)p^{\rm eq}(\e_0)
\ee
As in I, a Gausian approximation for the correlations of the dynamical variables $M(\e)$ will be used,
\be\label{MkMl.mit}
\lg M(\e)\rg=0
\quad\mbox{and}\quad
\lg M(\e)M(\e_0)\rg=\d(\e-\e_0)\lg M(\e)^2\rg
\ee
yielding $C_M(t,t_0)=\int\!d\e\lg M(\e)^2\rg G(\e,t-t_0|\e)p^{\rm eq}(\e)$.
For the trap model, the probability that the system returns to a trap that has been occupied earlier vanishes.
If there are $N$ traps, this means that one has $G(\e,t|\e_0)=\d(\e-\e_0)e^{-\k(\e)t}+{\cal O}(1/N)$.
One thus finds
\be\label{Pi.t}
\Pi_2(t)=\int\!d\e\lg M(\e)^2\rg e^{-\k(\e)t}p^{\rm eq}(\e)
\ee
The most obvious interpretation of this function is obtained for $\lg M(\e)^2\rg=1$, in which case $\Pi_2(t)$ is directly related to a dynamic structure factor at very large values for the scattering vector\cite{MB96}.
This means that every movement completely decorrelates the variable. 
A similar function also is observed in NMR when stimulated echo techniques are used\cite{G23} with a high spatial resolution, i.e. a large value of the evolution time $t_p$.
In that context, $\Pi_2(t)$ has been termed the angular jump function $F_2(t)$\cite{G42}.

Next, the four-time correlation function (4t-CF) occuring in eq.(\ref{R3.FDT}) will be calculated,
\be\label{C4.allg}
C_4(t,t_1,t_2,t_3)=\lg M(t)M(t_1)M(t_2)M(t_3)\rg
\ee
In order to proceed, I will again use the Gaussian factorization approximation
\Be\label{Mh4.mit.Gauss}
\lg M(\e_1)M(\e_2)M(\e_3)M(\e_4)\rg
&&\hspace{-0.6cm}=
\d(\e_1-\e_2)\d(\e_3-\e_4)\lg M(\e_1)^2\rg\lg M(\e_3)^2\rg
\nonumber\\
&&\hspace{-0.6cm}+\;
\d(\e_1-\e_3)\d(\e_2-\e_4)\lg M(\e_1)^2\rg\lg M(\e_2)^2\rg
\\
&&\hspace{-0.6cm}+\;
\d(\e_1-\e_4)\d(\e_2-\e_3)\lg M(\e_1)^2\rg\lg M(\e_2)^2\rg
\nonumber
\Ee
For the trap model, this yields
\be\label{C4.Pi4.Psi2}
C_4(t,t_1,t_2,t_3)=\Pi_4(t,t_1,t_2,t_3)+2\Psi_2(t-t_3)
\ee
where I defined 
\be\label{Psi.def}
\Psi_2(t)=\int\!d\e\lg M(\e)^2\rg^2 e^{-\k(\e)t}p^{\rm eq}(\e)
\ee
and the specific 4t-CF
\be\label{P4.def}
\Pi_4(t_0,t_1,t_2,t_3)
=\int\!d\e\!\int\!d\e'\lg M(\e)^2\rg\lg M(\e')^2\rg e^{-\k(\e)t_{01}}G(\e,t_{12}|\e')e^{-\k(\e')t_{23}}
   p^{\rm eq}(\e')
\ee
In this last expression, I used the abbreviation $t_{kl}=(t_k-t_l)$ to indicate that only the time intervals are relevant,
\be\label{Pi4.tdiffs}
\Pi_4(t_0,t_1,t_2,t_3)=\Pi_4(t_{01},t_{12},t_{23})
\ee
The special dynamic character of the trap model becomes particularly apparent if one uses energy-independent variables, $\lg M(\e)^2\rg=1$.
Setting $t_{01}=t_{23}=t_F$ and $t_{12}=t_{req}$, one recovers the 4t-function $F_4(t_F,t_{req})$
as it results in the framework of a free-energy landscape model for glassy relaxation, if the assumption of large 
$t_p$ is used, cf. eq.(16) in ref.\cite{G17},
\be\label{F4.def}
F_4(t_F,t_{req})
=\int\!d\e\!\int\!d\e' e^{-\k(\e)t_F}G(\e,t_{req}|\e')e^{-\k(\e')t_F}p^{\rm eq}(\e')
\ee
The interpretation of this function is quite straightforward.
In the first time interval, one selects slow molecules, i.e. only those molecules which have not reoriented during $t_F$ (due to the factor $e^{-\k(\e')t_F}$ in eq.(\ref{F4.def})). 
In the 're-equilibration' period, the system evolves freely, i.e. transitions among the different traps can take place (factor $G(\e,t_{req}|\e')$) and finally one only detects those molecules that still are slow 
($e^{-\k(\e)t_F}$). 
The fraction of selected molecules is determined by the limit of long $t_{req}$, denoted by $F_4(t_F,\infty)$.
Similar arguments can be set up for another prominent 4t-CF, which has been denoted 
$G_4(t_F,t_{req},t)\sim \Pi_4(t_F,t_{req},t)$. This function is not designed to detect the 'lifetime' of the dynamic heterogeneities, but the rotational decorrelation of a selected subensemble\cite{G42}.

\begin{figure}[h!]
\centering
\includegraphics[width=8.25cm]{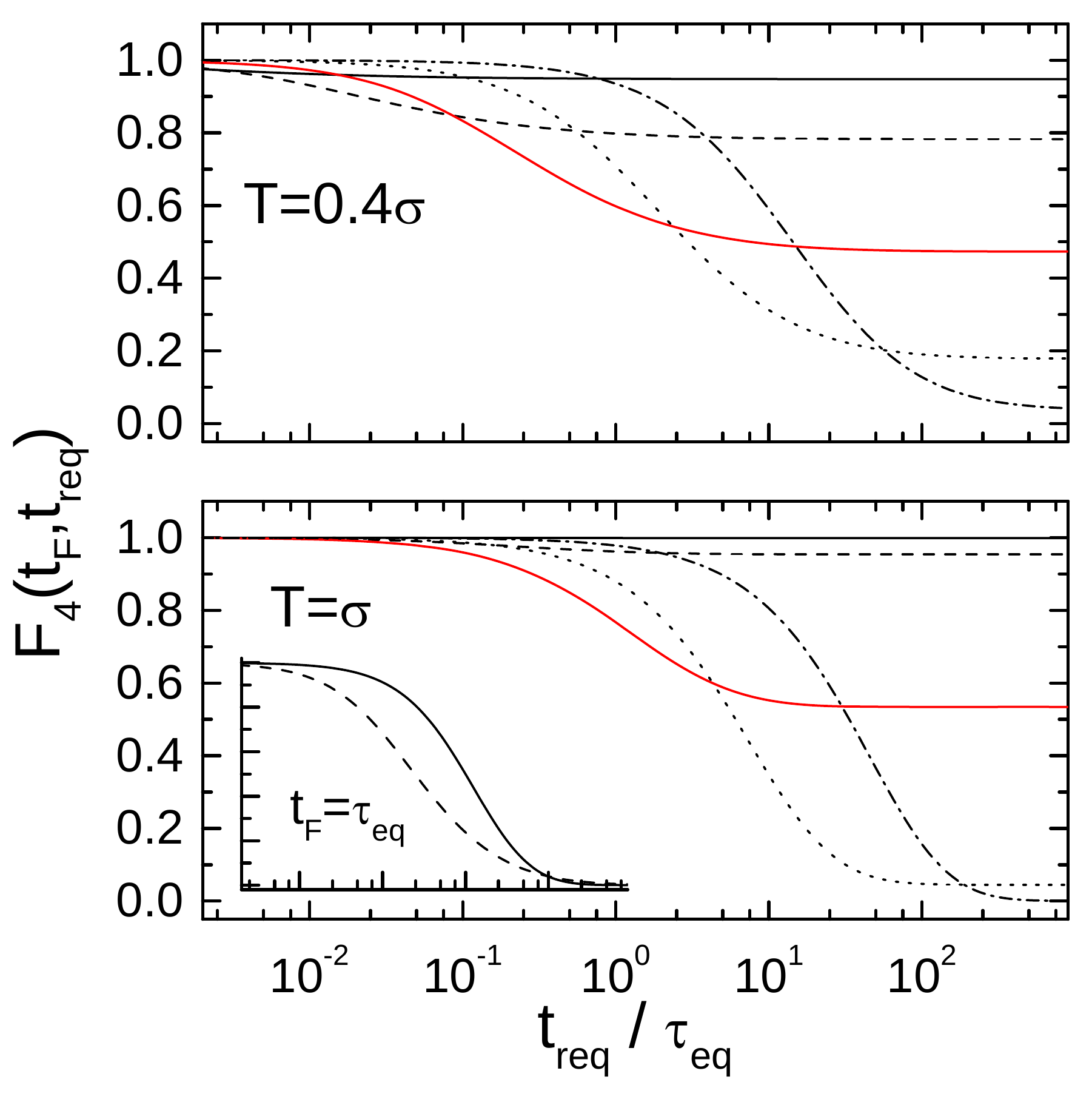}
\hspace{0.5cm}
\includegraphics[width=8.25cm]{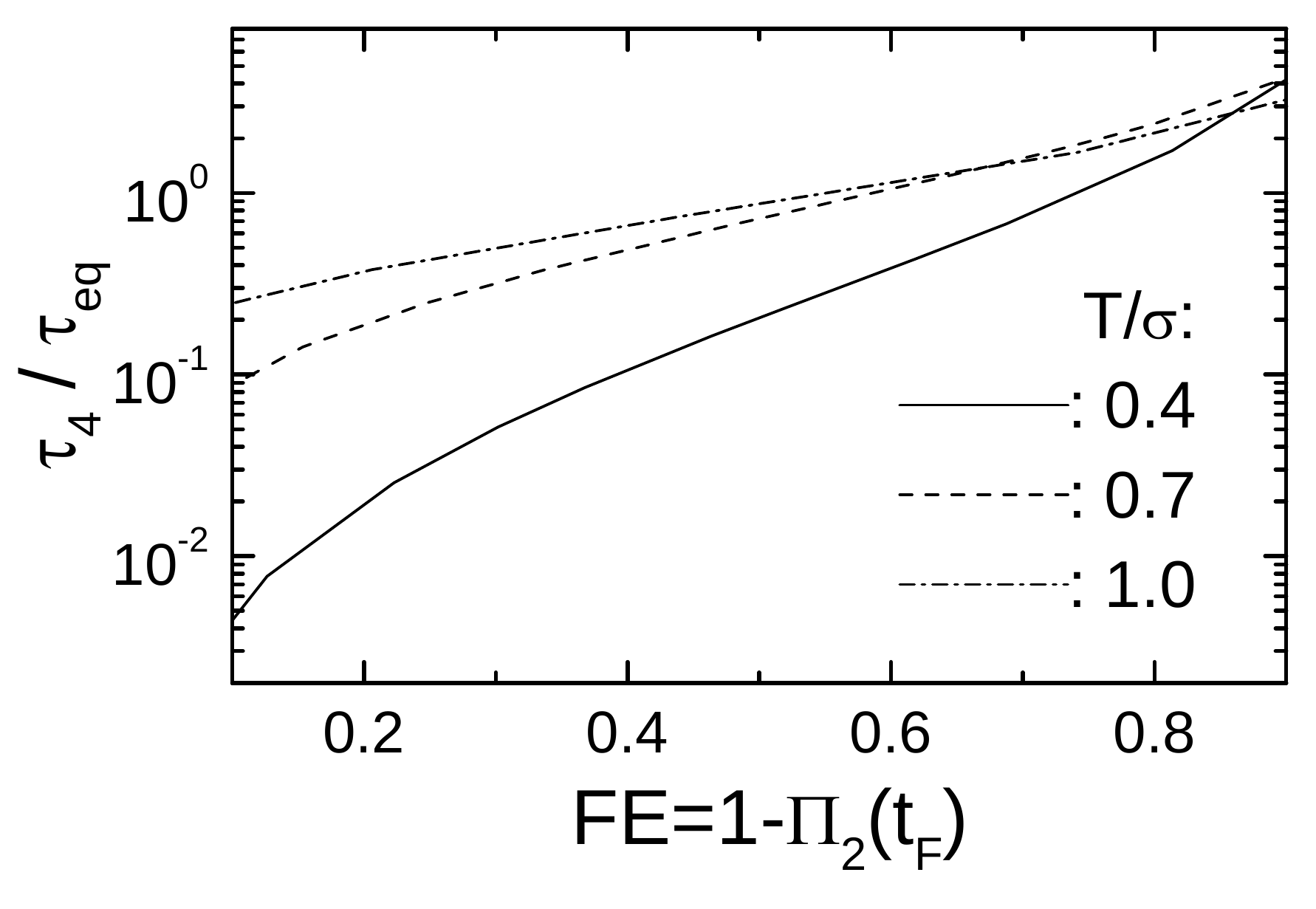}
\vspace{-0.5cm}
\caption{{\bf a}:(left) $F_4(t_F,t_{req})$ as a function of $t_{req}$ for temperatures $T/\s=0.4$ and $T/\s=1.0$. 
The filter times have been chosen as $t_F=10^{-2}$ (full lines), $t_F=10^{-1}$ (dashed lines), $t_F=1$ (red (grey) lines), $t_F=10$ (dotted lines) and $t_F=100$ (dot-dashed lines). All times are measured in units of $\t_{\rm eq}$.
The inset in the lower panel shows $F_4(t_F,t_{req})-F_4(t_F,\infty)/(1-F_4(t_F,\infty))$ for $t_F=1$ and 
$T=0.4\s$ (dashed line) and $T=\s$ (full line).
{\bf b}: (right) $\t_4/\t_{\rm eq}$ as a function of the so-called filter efficiency, FE\cite{G23}, where $\t_4$ denotes the $1/e$-decay time of $F_4(t_F,t_{req})$.
}
\label{Plot1}
\end{figure}

It is well known that the 2t-CF $\Pi_2(t)$ does not obey time-temperature superposition and the same holds for the function $F_4$, as can be observed in Fig.\ref{Plot1}. 
In Fig.\ref{Plot1}a, $F_4(t_F,t_{req})$ is shown for two temperatures as a function of the re-equilibration time.
Here, $\t_{\rm eq}=\k_\infty^{-1}e^{(3/2)\b^2\s^2}$ is the integral relaxation time of $\Pi_2(t)$ for $n=0$,
$\t_{\rm eq}=\int\!d\e \k(\e)^{-1}p^{\rm eq}(\e)$.
Note that $F_4(t_F,t_{req})$ decreases more exponentially at higher temperature, as shown for $t_F=1$ in the inset of Fig.\ref{Plot1}a.
This is similar to the 2t-CF and is a manifestation of the failure of time-temperature superposition in the trap model with a Gaussian DOS.
Fig.\ref{Plot1}b shows the time scale of the re-equilibration, denoted by $\t_4$. 
This time scale $\t_4$ is an increasing function of the filter efficiency FE\cite{G23}. 
The steepness in this plot simply reflects the different width of the underlying relaxation time distribution. 
All features are completely compatible with the fact that a given relaxation rate looses its memory on the time scale of the relaxation itself\cite{Heuer:1997}. 
Of course, for the trap model there is only one time scale for relaxation and thus this behavior is to be expected.

In a next step, consider an energy-dependent variable.
As in I, I choose the following form\cite{FS02}:
\be\label{Mh2.mit.n}
\lg M(\e)^2\rg=e^{-n\b\e}
\ee
with variable $n$ and where the static value of $M^2$ has been set to unity.
This choice leaves the 2t-CFs unaffected, because of 
\[
\int\!d\e p(\e)^{\rm eq}e^{-n\b\e}e^{-\k(\e)t}
=e^{{n(n+2)\over2}\b^2\s^2}\int\!d\e p(\e)^{\rm eq}e^{-\k(\e)t}
\]
with $t_n=t\cdot e^{-n\b^2\s^2}$.
Thus, only the time scale and the amplitude of $\Pi_2$ and $\Psi_2$ are changed, cf. also the discussion of the linear susceptibility in I.

However, the interpretation of $F_4(t_F,t_{req})$ in terms of selected subensembles can be flawed.
If energy-dependent variables are considered, the factor $\lg M(\e')^2\rg e^{-\k(\e')t_F}$ determines the dependence of the corresponding function $\Pi_4(t_F,t_{req},t_F)$ on the 'selection' period, cf. eq.(\ref{P4.def}).
If for instance, $n=1$ in eq.(\ref{Mh2.mit.n}), then sub-ensembles with large rates ($e^{\b\e}\gg1$) have very little weight in the selection ($e^{-\b\e}\ll1$). 
Thus, while for $n=0$, the plateau-value $F_4(t_F,\infty)$ is a monotonously decaying function of $t_F$, this must not hold in general.
\begin{figure}[h!]
\centering
\includegraphics[width=8.0cm]{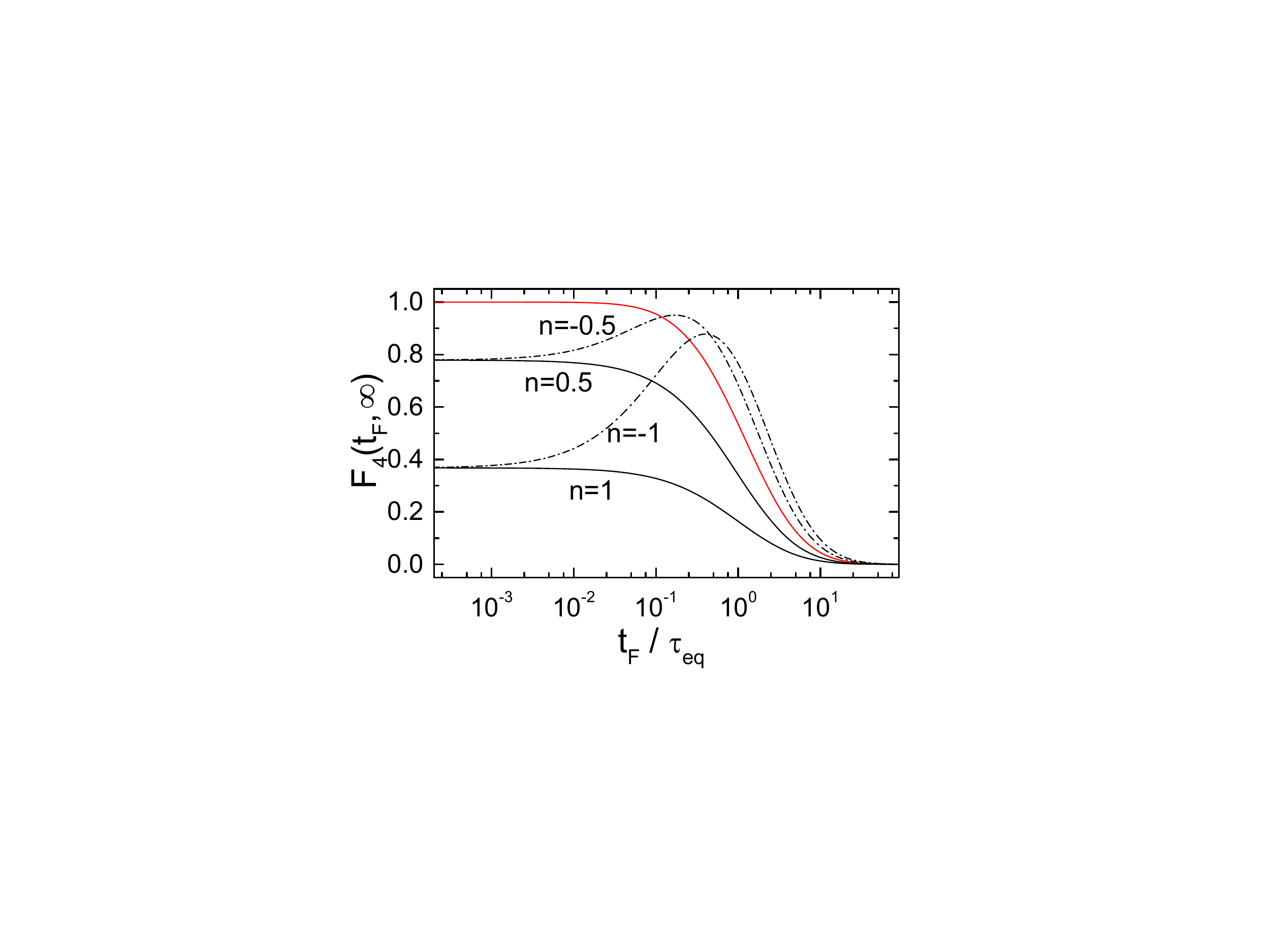}
\vspace{-0.52cm}
\caption{The plateau value $F_4(t_F,\infty)$ as a function of the filter time $t_F$ vor various different energy-dependencies of the dynamical variable, i.e. different values of $n$, as indicated.
The red (grey) line represents $n=0$.
}
\label{Plot2}
\end{figure}
In Fig.\ref{Plot2}, this plateau-value is shown for $T=\s$ and various values for $n$.
It is obvious, that for $n<0$ the fraction of selected molecules apparently first increases and decreases only for long $t_F$. 
This fact shows, that the character of the 4t-CFs can change dramatically for different variables, even though the 2t-CFs are unaffected by the choice of $n$.

Note that in the context of the four-dimensional NMR experiments, only $n=0$ is relevant and there is no ambuigity.
In general, however, the 4t-CFs cannot be interpreted in the simple picture and also the fact that there is no return probability in the trap model does not allow a simple interpretation.
This situation is very similar to the findings for the response, where the linear response does not change its features with $n$ but the cubic response strongly depends on the choice of $n$.
\section*{III. Nonlinear response functions from 'quasi-FDT'}
As mentioned above, in ref.\cite{Bouchaud05} it was argued that the most important behavior of the cubic response function is captured by eq.(\ref{R3.FDT}).
Given the 4t-CF, we are in the position to compare the results of this 'quasi-FDT' relation to the exact response for the same model.

Before discussing the results of model calculations, the most prominent experimental findings are summarized as already indicated in the Introduction.
One finds that there is a maximum in the one-$\om$ ($\a=1$) or three-$\om$ ($\a=3$) component of the modulus of $\chi_3(\om)$, 
\be\label{X.alfa.def}
X_\a(\om,T)={T\over(\D\chi_1)^2}|\chi_3^{(\a)}(\om)|\quad,\quad\a=1,\;3
\ee
with $\D\chi_1$ denoting the static linear susceptibility (which in dielectric spectroscopy corresponds to $\D\e$).
This maximum is a monotonously decaying function of temperature and its value has been related to the number of correlated particles or a length scale associated with the primary relaxation of the system\cite{CrausteThibierge10,Brun11}.

The findings for $\chi_3$ as obtained in I are summarized as follows.
As for a general Markovian dynamics that follows a ME, the response for the trap model depends on the details of the coupling of the external field to the transition rates in the sense that it is not irrelevant whether the field is coupled to the initial or the final state of a transition from one trap to the other. 
This dependence, however, is weak and does not alter the qualitative behavior of $X_\a(\om)$.
Much more important is the mentioned dependence of the cubic susceptibility on the dynamical variable chosen, i.e. on the value of $n$.
For $n=0$, i.e. variables that are independent of the trap-energy, one finds a maximum in $X_\a(\om)$.
In contrast to the experimental findings, however, this maximum $X_\a^{(\rm max)}=X_\a(\om_{\rm max})$ increases with increasing temperature.
For $n=1$, one finds 'trivial' behavior, i.e. a decay from a finite low-frequency value to a vanishing susceptibility at high frequency, for low temperatures. 
At some finite temperature a peak evolves in $X_\a(\om)$ with a more or less temperature-independent height.
For $n=-1$, one finds a maximum at all temperatures and one observes that $X_\a^{(\rm max)}$ decreases with increasing temperature, reminiscent to what is observed experimentally.

\begin{figure}[h!]
\centering
\includegraphics[width=8.0cm]{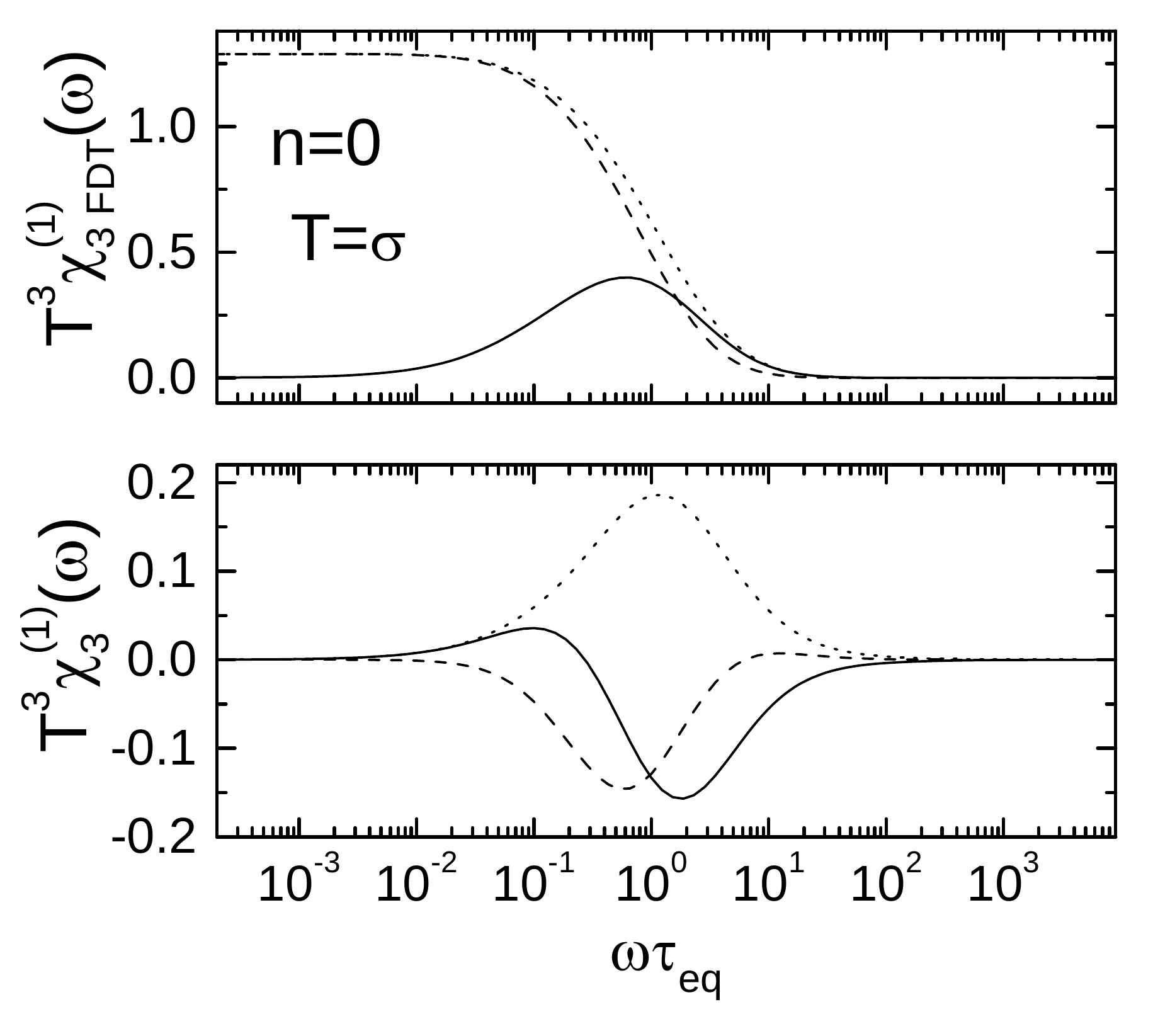}
\vspace{-0.5cm}
\caption{Cubic susceptibility $\chi_3^{(1)}(\om)$ according the the quasi-FDT relation, eq.(\ref{R3.FDT}) (upper panel) and according to the exact calculation (lower panel).
Full lines represent the imaginary part, dashed lines the real part of $\chi_3^{(1)}(\om)$ and the dotted lines are $X_1(\om)$ using eq.(\ref{X.alfa.def}). 
}
\label{Plot3}
\end{figure}

\begin{figure}[h!]
\centering
\includegraphics[width=8.0cm]{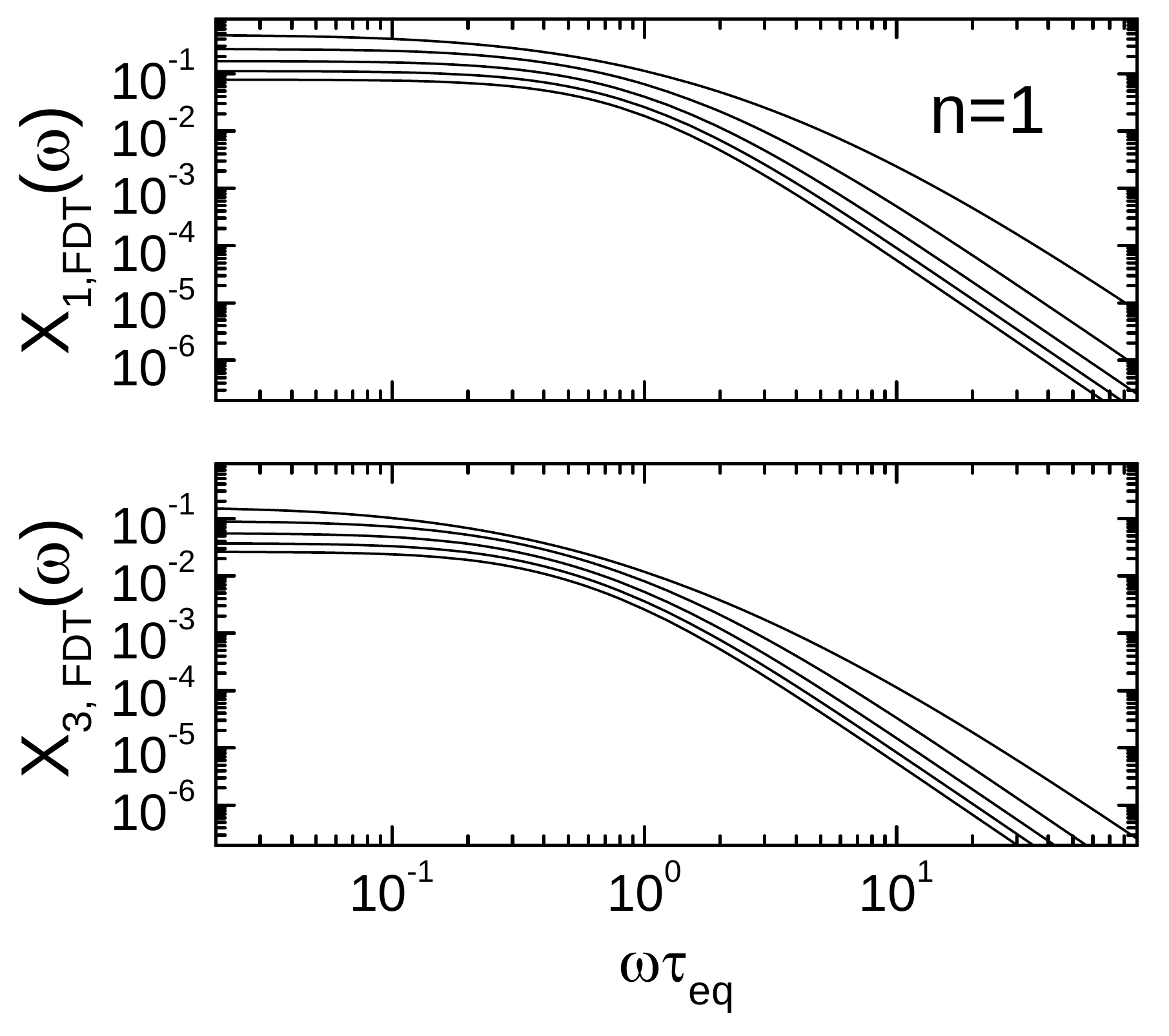}
\vspace{-0.5cm}
\caption{$X_{\a,FDT}(\om)$ as a function of frequency for various temperatures, $T/\s=1,\;1.5,\;2,\;2.5,\;3$ from top to bottom.
}
\label{Plot4}
\end{figure}

\begin{figure}[h!]
\centering
\includegraphics[width=8.0cm]{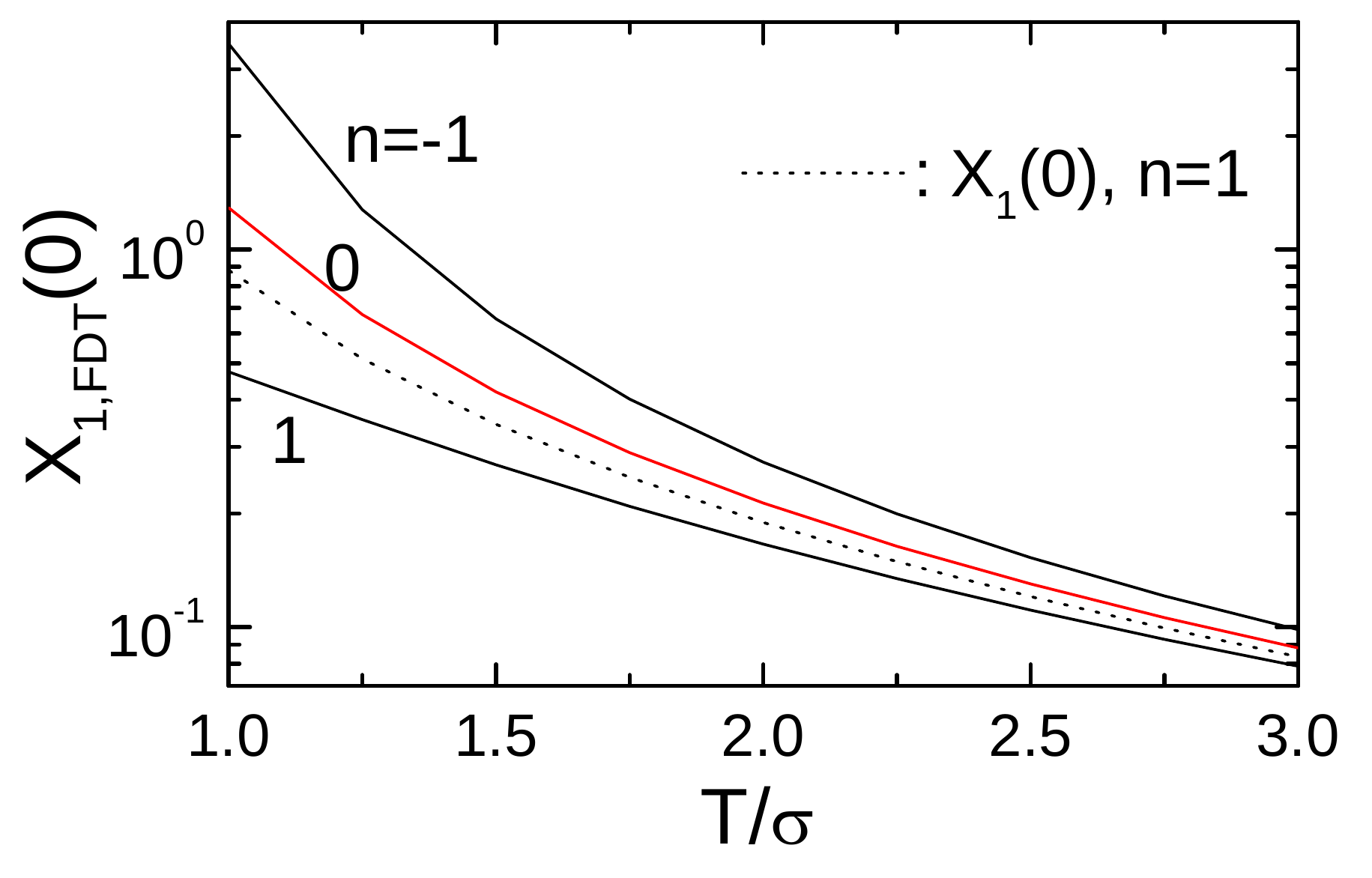}
\vspace{-0.5cm}
\caption{$X_{\a,FDT}(0)$ as a function of temperatures for different values of $n$.
The red (grey) line is for $n=0$.
For comparison, for $n=1$, the exact result also is shosn as dotted line.
}
\label{Plot5}
\end{figure}

Now, eq.(\ref{R3.FDT}) will be used to calculate the cubic response and the results are compared to the exact results as obtained in I.
According to eqns.(\ref{C4.allg}) and (\ref{C4.Pi4.Psi2}), one has
\[
R_{3,FDT}(t_0,t_1,t_2,t_3)=\b^3{d^3\over dt_1dt_2dt_3}C_4(t_0,t_1,t_2,t_3)=
\b^3{d^3\over dt_1dt_2dt_3}\Pi_4(t_0,t_1,t_2,t_3)
\]
The structure of $\Pi_4$ given in eq.(\ref{P4.def}) shows that there is a constant long-time level with regard to the second time interval $t_{12}$. 
This is because $G(\e,t|\e')$ tends towards $p^{\rm eq}(\e)$ at long times.
One can thus write $G(\e,t|\e')=p^{\rm eq}(\e)+\d G(\e,t|\e')$ where $\d G(\e,t|\e')$ decays to zero for long times.
This gives rise to the decomposition 
$\Pi_4(t_{01},t_{12},t_{23})=\Pi_4^{\rm eq}(t_{01},t_{23})+\d\Pi_4(t_{01},t_{12},t_{23})$.
For the response it is sufficient to consider ${d^3\over dt_1dt_2dt_3}\d\Pi_4(t_{01},t_{12},t_{23})$.
The general expression for the cubic susceptibility is given in the Appendix and the steady state result for an external field of the form $H(t)=H_0\cos{(\om t)}$ can be written as:
\be\label{Chi3.FDT}
\chi_{3,FDT}(t)={H_0\over2}\left[\sum_{\a=1,3}e^{-i\a\om t}\chi_{3,FDT}^{(\a)}(\om)+c.c\right]
\ee
where $H_0$ is the amplitude of the applied field and the $\chi_{3,FDT}^{(\a)}(\om)$ denote the corresponding frequency-components.
For a direct comparison, in Fig.\ref{Plot3} the cubic susceptibility is shown for $n=0$.
It is evident, that $X_{1,FDT}(\om)$ shows trivial behavior, whereas $X_{1}(\om)$ shows a peak.
From the imaginary and the real part of the susceptibilities it is apparent that the dispersive behavior in both cases is observed on the time scale of the relaxation time $\t_{\rm eq}$.
The main difference lies in the zero-frequency limit.
As an example, in Fig.\ref{Plot4}, $X_{\a,FDT}(\om)$ is shown for $n=1$ and for different temperatures and only 'trivial' behavior is found.
This is at variance with $X_{\a}(\om)$, where one observes trivial behavior at low temperatures and a peak at higher temperatures, cf. Figs.7 of I.
This trivial behavior of $X_{\a,FDT}(\om)$ is found independent of the value of $n$ and of temperature.
As already mentioned, the low-frequency limit $X_{\a,FDT}(0)$ is finite in all cases, as exemplified in 
Fig.\ref{Plot5}.
Note that for $n=0$ and for $n=-1$, the exact calculation gives $X_{\a}(0)=0$. 
For $n=1$, the finite value of $X_{\a}(0)$ does give rise to trivial behavior of $X_{\a}(\om)$ only at low temperature, but due to its decrease as a function of temperature, at higher temperatures a peak is observed.

These results show that when the dependence of the cubic response on the dynamical variable is discussed it is  important to note that the occurence of a peak in $X_\a(\om,T)$ strongly depends on the low-frequency limit 
$X_\a(0,T)$. 
Thus, while eq.(\ref{R3.FDT}) gives rise to a response which resembles the dynamic behavior of the exact response, for the trap model the results using this 'quasi-FDT' relation give the wrong values for static response with the consequence that only trivial behavior of $X_{\a,FDT}(\om)$ is observed.
These observation hold for $\a=1$ and $\a=3$ and as in the exact calculation one has
$X_{1,FDT}(0)=3X_{3,FDT}(0)$ for the low-frequency limit.

\section*{V. Conclusions}
I have computed the four-time correlation function for a trap model with a Gaussian density of states for
energy-dependent dynamical variables using a Gaussian factorization approximation. 
Due to the fact that in this model the probability of returning to the original trap vanishes, this 4t-CF for energy-independent variables strongly resembles the 4t-function that has been considered in many NMR investigations of the life time of the dynamic heterogeneities in supercooled liquids.
If, however, one allows for an energy-dependence of the dynamical variables, the interpretation of the corresonding 4t-CF no longer is straightforward because subensembles that relax at different rates have different weight in the average.
The specific choice made in eq.(\ref{Mh2.mit.n}), $\lg M(\e)^2\rg=\exp{(-n\b\e)}$, means that for $n=1$ slow subensembles have large weight while for $n=-1$ the fast subensembles dominate.
It is important to note that this different weighting does not affect the 2t-CFs. 
There, only the overall time scale and the amplitude depend on the value of $n$.
In the 4t-CF, the different weights prevent the characterization of the time intervals as filter interval and re-equilibration interval. 
As for many other properties of the energy-landscape of glasses, the Gaussian trap model captures the main features observed in four-dimensional NMR, particularly the fact that there is only a single time scale in the relaxation.
This means that the 'exchange-time' is on the order of the relaxation time of the primary relaxation or the rate memory parameter is close to unity.

When the cubic response is considered, in general there is no obvious relation to the 4t-CF. 
However, it has been argued that a 'quasi-FDT' relation, eq.(\ref{R3.FDT}), relating the cubic response to a triple time-derivative of the 4t-CF captures the important physics\cite{Bouchaud05}.
I have used this recipe in order to compute the cubic susceptibility and compare it to the exact results obtained in I.
One finds that the response takes place on the same time scale as in the exact calculation. 
This means that the approximation basically does not change the dispersive behavior of the response.
This, however, is to be expected because in the trap model there is only a single time scale.
Regarding the modulus of the cubic response, the exact results differ strongly from the approximative ones based on eq.(\ref{R3.FDT}).
One finds that for all choices of the dynamical variables and independent of temperature the quantity
$X_\a(\om,T)$ defined in eq.(\ref{X.alfa.def}) shows 'trivial' behavior, i.e. a monotonous decay from a finite low-frequency value to zero at high frequencies.
In the exact calculations, for all values of $n$ a peak is observed in $X_\a(\om,T)$, albeit not necessarily for all temperatures.
Thus, for the trap model the dynamic behavior of the cubic response appears to be described reasonably by 
eq.(\ref{R3.FDT}) but the static susceptibility is wrong.
This makes the interpretation very difficult since it changes the character of $X_\a(\om,T)$ completely.

More generally, due to the dissipative nature of the response on the relaxation time scale, one expects a strong variation of the susceptibility in the corresponding frequency range.
Only if the peak resulting in the modulus from this dispersion is not too small compared to the low-frequency limit, a hump located at a frequency on the order of the inverse relaxation time should develop.
This is the reason why in the Debye model no hump is observed due to a rather large zero-frequency value of the response\cite{Dejardin00}.
The same argument applies to the ADWP-model for dipole reorientations in the case of vanishing asymmetry.
For finite asymmetry, in this model, a hump is observed in a certain temperature range around the temperature at which $X_\a(0,T)$ vanishes, cf. the discussion in I. 
The main difference between the ADWP-model and the model of Brownian rotation lies in the fact that in the former model $X_\a(0,T)$ depends on temperature for finite asymmetry while for the latter model it is temperature-independent.
Consequently, these two models mainly show trivial behavior.
The interplay between the static susceptibility and the intensity of the dispersion due to the dynamics
determines the behavior of $X_\a(\om,T)$ and in particular the possible occurence of a hump-like structure.
Consequently, every approximation made in the computation of the response must yield the correct behavior at least qualitatively.

In summary, I have shown that the 4t-CFs of the Gaussian trap model show the features of dynamic heterogeneities that have been observed in four-dimensional NMR experiments.
The use of the 4t-CFs in the calculation of the cubic response is problematic for this model because it does not give a meaningful approximation to the exact response.
It is important to note that this finding cannot be generalized to other models in an obvious way and therefore further calculations are required.
In particular, it is important to perform calculations for models that follow a Langevin dynamics, because originally eq.(\ref{R3.FDT}) was derived for this case.
\subsection*{Acknowledgment}
I thank Roland B\"ohmer, Gerald Hinze, and Jeppe Dyre for fruitful discussions.
%
\begin{appendix}
\subsection*{Appendix: Cubic 'quasi-FDT' response for the trap model}
\setcounter{equation}{0}
\renewcommand{\theequation}{A.\arabic{equation}}
The general expression for the cubic response can be written as (cf. the Appendix of I):
\be\label{Chi3.FDT.allg}
\chi_{3,FDT}(t)=
\int_{t_0}^t\!dt_1H(t_1)\int_{t_0}^{t_1}\!dt_2H(t_2)\int_{t_0}^{t_2}\!dt_3H(t_3)
R_{3,FDT}(t,t_1,t_2,t_3)
\ee
In order to compute the Fourier-components, I start from the discrete version of the expression for 
$\Pi_4(t_{01},t_{12},t_{23})$, eq.(\ref{P4.def}),
\[
\Pi_4(t_{01},t_{12},t_{23})=
\sum_{k,l}\lg M_k^2\rg\lg M_k^2\rg e^{-\k_kt_{01}}G_{kl}(t_{12})e^{-\k_lt_{23}}p_l^{\rm eq}
\]
with $G_{kl}(t)=G(\e_k,t|\e_l)$, $\k_k=\k(\e_k)$ and $\lg M_k^2\rg=\lg M(\e_k)^2\rg$.
The Greens function can be expressed in terms of the eigenvalues $\l_m$ and the eigenvectors $S_{km}$ of the symmetrized matrix of transition rates\cite{vkamp81},
\be\label{Gkl.S.lam}
G_{kl}(t)=\sqrt{{p_k^{\rm eq}\over p_l^{\rm eq}}}\sum_{m=0}^N S_{km}S_{lm}e^{-\G_mt}
\ee
Here, the eigenvalues $\l_m=-\G_m$ all are negative (and the $\G_m$ are relaxation rates) and for the $m=0$ one has 
$\l_0=0$ and $S_{k0}=\sqrt{p_k^{\rm eq}}$.
Consequently, this yields the long-time limit, $G_{kl}(t\to\infty)=p_k^{\rm eq}$.
Thus, the relevant function, $\d\Pi_4(t_{01},t_{12},t_{23})$, can be expressed as a sum including the terms with $m>0$.
Performing the time derivatives and the Fouriertransform yields in the steady state
\be\label{Chi3.FDT.result}
\chi_{3,FDT}^{(\a)}(\om)
={(-1)\over4}\b^3\sum_{k,l}\sum_{m>0}\lg M_k^2\rg\lg M_k^2\rg S_{km}S_{lm}(\G_m-\k_k)(\G_m-\k_l)
\k_l\sqrt{p_k^{\rm eq}p_l^{\rm eq}}{\cal F}_{kml}^{(\a)}(\om)
\ee
with the spectral functions
\Be\label{Sklm.n}
{\rm Re}({\cal F}_{kml}^{(1)}(\om))
&&\hspace{-0.6cm}=
{3\k_k\G_m^2\k_l+\om^2(8\k_k\k_l-2\G_m(\k_k+\k_l)-\G_m^2)
                    \over\G_m(\k_k^2+\om^2)(\k_l^2+\om^2)(\G_m^2+4\om^2)}
\nonumber\\                
{\rm Im}({\cal F}_{kml}^{(1)}(\om))
&&\hspace{-0.6cm}=                
\om{\k_k\G_m^2+2\k_k\G_m\k_l+3\G_m^2\k_l+2\om^2(4\k_l-\G_m)
                    \over\G_m(\k_k^2+\om^2)(\k_l^2+\om^2)(\G_m^2+4\om^2)}
\\                
{\rm Re}({\cal F}_{kml}^{(3)}(\om))
&&\hspace{-0.6cm}=
{\k_k\G_m\k_l-\om^2(2\k_k+3\G_m+6\k_l)
                 \over(\k_l^2+\om^2)(\G_m^2+4\om^2)(\k_k^2+9\om^2)}
\nonumber\\                
{\rm Im}({\cal F}_{kml}^{(3)}(\om))
&&\hspace{-0.6cm}=                
\om{\G_m(\k_k+3\k_l)+2\k_k\k_l-6\om^2
                    \over(\k_l^2+\om^2)(\G_m^2+4\om^2)(\k_k^2+9\om^2)}
\nonumber            
\Ee                
\end{appendix}
\newpage

%
\end{document}